\documentclass[longauth]{aa}  
\usepackage{graphicx}
\usepackage{txfonts}
\usepackage{multirow}
\usepackage{multicol}
\usepackage[flushleft]{threeparttable}
\usepackage{hyperref}
\hypersetup{colorlinks=true, citecolor=blue, linkcolor=blue}
\usepackage{mathtools}

\newcommand{\mbh}{{\mbox{$M_\mathrm{BH}$}}}

\newcommand{\ha}{{\mbox{H$\alpha$}}}
\newcommand{\hb}{{\mbox{H$\beta$}}}
\newcommand{\brg}{{\mbox{Br$\gamma$}}}
\newcommand{\paa}{{\mbox{Pa$\alpha$}}}

\newcommand{\feii}{Fe{\sevenrm\,II}}

\newcommand{\OIIIc}{[O{\sevenrm\,III}]\,$\lambda\lambda$4959,5007}

\newcommand{\lopt}{\mbox{$\lambda L_\lambda$ (5100 \AA)}}
\newcommand{\Rd}{\mbox{$R_\mathrm{d}$}}
\newcommand{\Rb}{\mbox{$R_\mathrm{BLR}$}}
\newcommand{\rEdd}{\mbox{$\lambda_\mathrm{Edd}$}}
\newcommand{\rl}{\mbox{$R$--$L$}}
\newcommand{\nodata}{{...}}
\newcommand{\mcl}{\multicolumn}

\newcommand{\um}{{\mbox{$\mu$m}}}
\newcommand{\kms}{{\mbox{$\mathrm{km\,s^{-1}}$}}}
\newcommand{\ergs}{{\mbox{$\mathrm{erg\,s^{-1}}$}}}
\newcommand{\msun}{{\mbox{$M_\odot$}}}

\font\sevenrm=cmr7 scaled 1000

\begin{document}

\titlerunning{Supermassive black hole mass from interferometric observation of dust continuum}
\title{Towards measuring supermassive black hole masses with interferometric observations of the dust continuum}

\authorrunning{GRAVITY Collaboration}
\author{GRAVITY Collaboration\thanks{GRAVITY is developed in a collaboration by 
the Max Planck Institute for Extraterrestrial Physics, LESIA of Observatoire 
de Paris/Universit\'e PSL/CNRS/Sorbonne Universit\'e/Universit\'e de Paris and 
IPAG of Universit\'e Grenoble Alpes /CNRS, the Max Planck Institute for Astronomy, 
the University of Cologne, the CENTRA - Centro de Astrofisica e Gravita\c c\~ao, 
and the European Southern Observatory.
Corresponding authors: J.~Shangguan (shangguan@mpe.mpg.de) and Y.~Cao (ycao@mpe.mpg.de)}:
A.~Amorim\inst{15,17}
\and G.~Bourdarot\inst{1}
\and W.~Brandner\inst{18} 
\and Y.~Cao\inst{1} 
\and Y.~Cl\'enet\inst{2} 
\and R.~Davies\inst{1}
\and P.~T.~de~Zeeuw\inst{1,13} 
\and J.~Dexter\inst{20,1}
\and A.~Drescher\inst{1} 
\and A.~Eckart\inst{3,14} 
\and F.~Eisenhauer\inst{1} 
\and M.~Fabricius\inst{1}
\and N.~M.~F\"orster~Schreiber\inst{1} 
\and P.~J.~V.~Garcia\inst{11,16,17} 
\and R.~Genzel\inst{1,4} 
\and S.~Gillessen\inst{1} 
\and D.~Gratadour\inst{2,21} 
\and S.~H\"onig\inst{5}
\and M.~Kishimoto\inst{6} 
\and S.~Lacour\inst{2,12} 
\and D.~Lutz\inst{1} 
\and F.~Millour\inst{7}  
\and H.~Netzer\inst{8} 
\and T.~Ott\inst{1} 
\and T.~Paumard\inst{2} 
\and K.~Perraut\inst{9} 
\and G.~Perrin\inst{2}
\and B.~M.~Peterson\inst{22}
\and P.~O.~Petrucci\inst{9} 
\and O.~Pfuhl\inst{12}
\and M.~A.~Prieto\inst{19}  
\and D.~Rouan\inst{2}
\and D.~J.~D.~Santos\inst{1}
\and J.~Shangguan\inst{1}
\and T.~Shimizu\inst{1}
\and A.~Sternberg\inst{8,10} 
\and C.~Straubmeier\inst{3} 
\and E.~Sturm\inst{1} 
\and L.~J.~Tacconi\inst{1} 
\and K.~R.~W.~Tristram\inst{11}  
\and F.~Widmann\inst{1} 
\and J.~Woillez\inst{12}}

\institute{
Max Planck Institute for Extraterrestrial Physics (MPE), Giessenbachstr.1, 
85748 Garching, Germany
\and LESIA, Observatoire de Paris, Universit\'e PSL, CNRS, 
Sorbonne Universit\'e, Univ. Paris Diderot, Sorbonne Paris Cit\'e, 5 place Jules Janssen, 
92195 Meudon, France
\and I. Institute of Physics, University of Cologne, Z\"ulpicher Stra{\ss}e 77, 
50937 Cologne, Germany
\and Departments of Physics and Astronomy, Le Conte Hall, University of California, 
Berkeley, CA 94720, USA
\and Department of Physics and Astronomy, University of Southampton, Southampton, UK
\and Department of Physics, Kyoto Sangyo University, Kita-ku, Japan
\and Universit\'e C\^ote d'Azur, Observatoire de la C\^ote d'Azur, CNRS, 
Laboratoire Lagrange, Nice, France
\and School of Physics and Astronomy, Tel Aviv University, Tel Aviv 69978, Israel
\and Univ. Grenoble Alpes, CNRS, IPAG, 38000 Grenoble, France
\and Center for Computational Astrophysics, Flatiron Institute, 162 5th Ave., 
New York, NY 10010, USA
\and European Southern Observatory, Casilla 19001, Santiago 19, Chile
\and European Southern Observatory, Karl-Schwarzschild-Str. 2, 85748 Garching, Germany
\and Sterrewacht Leiden, Leiden University, Postbus 9513, 2300 RA Leiden, The Netherlands
\and Max Planck Institute for Radio Astronomy, Auf dem H\"ugel 69, 53121 Bonn, Germany
\and Universidade de Lisboa - Faculdade de Ci\^{e}ncias, Campo Grande, 
1749-016 Lisboa, Portugal
\and Faculdade de Engenharia, Universidade do Porto, rua Dr. Roberto Frias, 
4200-465 Porto, Portugal
\and CENTRA - Centro de Astrof\'isica e Gravita\c{c}\~{a}o, IST, Universidade de Lisboa, 
1049-001 Lisboa, Portugal
\and Max Planck Institute for Astronomy, K\"onigstuhl 17, 69117, Heidelberg, Germany
\and Instituto de Astrof\'isica de Canarias (IAC), E-38200 La Laguna, Tenerife, Spain
\and Department of Astrophysical \& Planetary Sciences, JILA, University of Colorado, 
Duane Physics Bldg., 2000 Colorado Ave, Boulder, CO 80309, USA
\and Research School of Astronomy and Astrophysics, Australian National University, 
Canberra, ACT 2611, Australia
\and Retired
}

\abstract{This work focuses on active galactic nuclei (AGNs), and the relation 
between the sizes of the hot dust continuum and the broad-line region (BLR). 
We find that the continuum size measured using optical/near-infrared 
interferometry (OI) is roughly twice that measured by reverberation mapping (RM). 
Both OI and RM continuum sizes show a tight relation with the \hb\ BLR 
size with only an intrinsic scatter of 0.25~dex.  The masses 
of supermassive black holes (BHs) can hence be simply derived from a dust size 
in combination with a broad line width and virial factor. Since the primary 
uncertainty of these BH masses comes from the virial factor, the accuracy of 
the continuum-based BH masses is close to those based on the RM measurement of 
the broad emission line. 
Moreover, the necessary continuum measurements can be obtained on a much shorter 
timescale than those required monitoring for RM, and are also more time efficient 
than those needed to resolve the BLR with OI.  The primary goal of this 
work is to demonstrate measuring the BH mass based on the dust continuum size 
with our first calibration of the \Rb--\Rd\ relation.  The current limitation 
and caveats are discussed in detail.  Future GRAVITY observations are expected 
to improve the continuum-based method and have the potential to measure BH 
masses for a large sample of AGNs in the low-redshift Universe.
} 
\keywords{galaxies: active -- galaxies: nuclei -- galaxies: Seyfert -- quasars: supermassive black holes}

\maketitle
%

\section{Introduction}

Measuring the mass of supermassive black holes (BHs) is challenging as this 
requires resolving stellar or gas dynamics inside the BH's sphere of 
influence 
\citep[e.g.][]{Thomas2004,Onken2014,Saglia2016,Hicks2008,Davis2014,Onishi2017,Boizelle2019}.
In active galactic nuclei (AGNs) with broad recombination lines, the reverberation mapping 
(RM) method has been developed to measure the size of the broad-line region 
(BLR) 
and hence lead to the measurement of the BH mass 
\citep{Blandford1982,Peterson1993,Peterson2004}.  
By monitoring the variability of the UV/optical continuum and an 
emission line, typically \hb, the BLR size can be obtained from the RM method. 
Assuming that the BLR is a virialized system, one can calculate the BH mass,
\begin{equation}\label{eq:vir}
\mbh = f \frac{\Rb (\Delta V)^2}{G},
\end{equation}
where \Rb\ is the BLR radius, $\Delta V$ is the velocity width of the broad 
emission line, $f$ is the corresponding virial factor, and $G$ is the 
gravitational constant.  The FWHM or 
second moment ($\sigma_\mathrm{line}$) of the broad line is usually used as $\Delta V$.  
The virial factor $f$ depends on the geometry, kinematics, 
and inclination of the BLR clouds 
and is likely different from object to object.  
For example, the virial factor is 0.75 assuming an isotropic velocity distribution of 
Keplerian motion \citep{Netzer1990}.  The mean virial factor, 
$\langle f \rangle$, can be calibrated using nearby AGNs assuming that the AGN 
and quiescent galaxies follow the same \mbh--$\sigma_*$ relation 
\citep[e.g.][]{Onken2004,Woo2010,Graham2011,Grier2013,Ho2014}.  
This method based on RM measurements has been used successfully for many years, 
despite potential biases and caveats in the virial factor calibration 
\citep{Shankar2019}.  
For a particular source, the uncertainty in $f$ may come from its 
unknown inclination angle and other effects such as the radiation pressure 
\citep{Collin2006,MejiaRestrepo2018}.  

Moreover, a scaling relation between the BLR radius and AGN luminosity was 
discovered from  RM measurements 
\citep[][and references therein]{Kaspi2000,DallaBonta2020}.  
This \rl\ relation enables one to estimate the BH mass only with the AGN 
luminosity and the FWHM of a broad emission line from single-epoch spectra 
\citep{Shen2013}.  Thanks to its simplicity, the single-epoch method has been 
widely used with different broad lines in the UV and optical, although 
the uncertainty is around 0.5~dex or above \citep[e.g.][]{Vestergaard2006}.
Other methods have been developed to estimate the BH mass and tested against 
the RM measured BH mass.  For example, the coronal line 
[Si~{\sevenrm VI}]1.963~\um can be used to estimate the BH mass with 
an uncertainty of about 0.5~dex \citep{Prieto2022}.

Recently, the BLRs of three AGNs have been spatially resolved by GRAVITY, a 
second-generation Very Large Telescope Interferometer (VLTI) instrument 
\citep{GC2018,GC2020iras,GC2021ngc3783}.  GRAVITY has greatly improved the 
sensitivity of earlier efforts and has been able to combine all four of the 8-m 
Unit Telescope (UT) beams to yield six simultaneous baselines \citep{GC2017FL}. 
With a few hours of on-source exposure, GRAVITY measures the differential phase 
signal of a broad emission line in the near-infrared (NIR) $K$-band, 
which reflects the offsets of the photocenters from the center of the continuum 
emission in each wavelength channel \citep{Petrov2001,Marconi2003}. 
The BLR size is then inferred from the differential phase data by fitting a dynamical BLR model such as the widely used \cite{Pancoast2014a} model. 
The GRAVITY-measured BLR size and BH mass are in good agreement with RM 
measurements \citep{GC2021ngc3783,GC2021dist}.  

GRAVITY can also resolve the size of the NIR continuum emission of 
the AGN, which comes from the thermal radiation of hot dust that is reprocessing 
the UV/optical continuum from the accretion disk 
\citep[e.g.][]{Rees1969,Barvainis1987}.  Spatial sizes of the dust continuum 
emission can be measured by both continuum RM 
\citep[e.g.][]{Clavel1989,Baribaud1992,Glass1992,Sitko1993,Minezaki2004,Suganuma2006,Koshida2014,
PozoNunez2014,PozoNunez2015,Mandal2018,Minezaki2019,Figaredo2020,Mandal2021a,Mandal2021b} 
as well as optical/NIR interferometry 
\citep[OI;][]{Swain2003,Wittkowski2004,Kishimoto2009,Kishimoto2011,GC2020cont,Leftley2021}.
Similar to the BLR size,  the dust continuum size also scales with the AGN 
luminosity $\propto L^{0.5}$ 
\citep{Suganuma2006,Kishimoto2011,Koshida2014,Minezaki2019,GC2020cont}.
Such a relation is expected if the dust temperature and the inner radius 
of the dust distribution are determined by radiation equilibrium and dust 
sublimation, respectively \citep{Barvainis1987,Kishimoto2007}.
The dust continuum RM radius is a factor of four or five larger than the BLR 
radius \citep{Koshida2014} and is consistent with BLR models that place
hot dust on the outskirts of the BLR \citep[e.g.][]{Wang2017,Baskin2018}.  
Moreover, RM-measured dust continuum sizes are systematically smaller than those 
measured from OI, likely because the RM size is weighted by the time lag over 
the emitting region while the OI size is weighted by the intensity of hot dust 
emission \citep{Koshida2014,Kishimoto2011,GC2020cont}.

GRAVITY can observe the dust continuum independently of the full 
spectroastrometric measurements and has demonstrated excellent efficiency (e.g. 
$\lesssim 1$~hour per source; \citealt{GC2020cont} and in preparation).
Establishing a link between the BLR and dust continuum size will enable BH mass 
estimations from these more accessible dust continuum observations. 
In this work, we investigate the correlation between BLR and dust 
continuum size in the context of estimating the BH mass.  The four methods discussed 
in this paper are:
\begin{enumerate}
\item Reverberation mapping of the broad emission line, where a sequence of 
measurements over months or years, yielding the time delay for variations in 
the broad-line emission, leads to an estimate \Rb; and hence, 
via Equation~(\ref{eq:vir}), the BH mass \citep{Peterson2014}.  
This method has enabled empirical calibration of 
a sample-average virial factor $\langle f \rangle$.  More recently, the velocity 
resolved RM data can constrain a BLR dynamical model and enable the estimation 
of $f$ for individual sources \citep[e.g.][]{Pancoast2014a,Pancoast2014b}.
\item The single-epoch method, where \Rb\ is estimated from a single measurement 
of the AGN luminosity via the relation between these quantities 
\citep{Shen2013,DallaBonta2020}.  This method relies on Equation~(\ref{eq:vir}) 
and the virial factor calibrated via RM.
\item Continuum size measurements, the method introduced in this paper, where 
\Rb\ is estimated from an interferometric measurement of \Rd.  
Like the single-epoch method, this also relies on Equation~(\ref{eq:vir}) and 
a pre-calibration of the virial factor.
\item Spectrally resolved differential phase measurements of the broad-line 
emission.  Using the interferometric data as constraints on a dynamical BLR 
model allows one to derive BH mass; and hence also infer a value of the virial 
factor for individual sources that is independent of RM \citep{GC2020iras}.
\end{enumerate}
We show that the \hb\ BLR size scales tightly with the dust continuum size, 
which allows us to estimate the BH mass from the dust continuum size with 
an uncertainty similar to the RM BLR method.
We discuss the prospects of this method for BH mass estimations in 
the low-redshift Universe, especially with the upgrade of GRAVITY in Section~\ref{sec:fut}.  
This work adopts the following parameters for a $\Lambda$CDM cosmology: 
$\Omega_m = 0.308$, $\Omega_\Lambda = 0.692$, and 
$H_{0}=67.8$~km s$^{-1}$~Mpc$^{-1}$ \citep{Planck2016AA}.

\section{Samples}
\label{sec:smp}

\begin{table*}
\caption{Physical properties of AGNs with dust continuum size measurements}
\renewcommand{\arraystretch}{1.2}
\scriptsize
\begin{center}
\begin{tabular}{l c r@{$\pm$}l c r@{$\pm$}l c r@{$\pm$}l r@{$\pm$}l c c r@{$\pm$}l r}
\hline\hline
            Name     & Redshift &  \mcl{2}{c}{\Rd\ (RM)} &    Ref. &  \mcl{2}{c}{\Rd\ (OI)} &    Ref. &     \mcl{2}{c}{\Rb} &    \mcl{2}{c}{FWHM} & $\log\,\lopt$ &     Ref. & \mcl{2}{c}{$\log\,\mbh$} &    $\log\,\rEdd$ \\       
                     &          &       \mcl{2}{c}{(ld)} &         &       \mcl{2}{c}{(ld)} &         &    \mcl{2}{c}{(ld)} &  \mcl{2}{c}{(\kms)} &       (\ergs) &          &      \mcl{2}{c}{(\msun)} &                  \\       
             (1)     &      (2) &        \mcl{2}{c}{(3)} &     (4) &        \mcl{2}{c}{(5)} &     (6) &     \mcl{2}{c}{(7)} &     \mcl{2}{c}{(8)} &           (9) &     (10) &         \mcl{2}{c}{(11)} & \mcl{1}{c}{(12)} \\\hline 
         Mrk 335     &   0.0258 &     167.5 &        6.0 &       1 &       185 &         49 &       2 &     14.0 &      4.0 &       1707 &     79 &          43.8 &        3 &         6.90 &      0.13 &          $-0.29$ \\       
         UGC 545     &   0.0612 &    \mcl{2}{c}{\nodata} & \nodata &       707 &         77 &       4 &     37.2 &      4.7 &       1131 &     37 &          44.5 &        5 &         6.97 &      0.06 &             0.39 \\       
     Mrk 590$^{a,b}$ &   0.0264 &      33.5 &        4.2 &       1 &    \mcl{2}{c}{\nodata} & \nodata &     25.6 &      5.9 &       2716 &    202 &          43.5 &        3 &         7.57 &      0.12 &          $-1.21$ \\       
          3C 120     &   0.0330 &      94.4 &        5.5 &       6 &       379 &         85 &       2 &     26.2 &      7.7 &       2472 &    729 &          44.0 &        3 &         7.49 &      0.29 &          $-0.64$ \\       
       H0507+164     &   0.0179 &        35 &         11 &       7 &    \mcl{2}{c}{\nodata} & \nodata &      3.0 &      1.2 &       4062 &    247 &          42.6 &        8 &         6.99 &      0.18 &          $-1.57$ \\       
         Akn 120     &   0.0327 &       138 &         18 &       1 &       387 &         77 &       4 &     39.5 &      8.2 &       6077 &    147 &          43.9 &        3 &         8.45 &      0.09 &          $-1.73$ \\       
   MCG+08-11-011     &   0.0205 &      72.7 &        1.6 &       1 &    \mcl{2}{c}{\nodata} & \nodata &     15.7 &      0.5 &       4139 &    207 &          43.3 &        3 &         7.72 &      0.05 &          $-1.54$ \\       
           Mrk 6$^a$ &   0.0195 &    \mcl{2}{c}{\nodata} & \nodata &       214 &         60 &       9 &     18.5 &      2.5 &       5457 &     16 &          43.6 &       10 &         8.03 &      0.06 &          $-1.55$ \\       
          Mrk 79     &   0.0222 &      67.7 &        4.8 &       1 &    \mcl{2}{c}{\nodata} & \nodata &     15.6 &      5.0 &       4793 &    145 &          43.7 &        3 &         7.84 &      0.14 &          $-1.31$ \\       
     PG 0844+349     &   0.0640 &        99 &         11 &       1 &    \mcl{2}{c}{\nodata} & \nodata &     32.3 &     13.6 &       2694 &     58 &          44.2 &        3 &         7.66 &      0.18 &          $-0.59$ \\       
         Mrk 110     &   0.0353 &     116.6 &        6.3 &       1 &    \mcl{2}{c}{\nodata} & \nodata &     25.6 &      8.1 &       1634 &     83 &          43.7 &        3 &         7.13 &      0.14 &          $-0.61$ \\       
     PG 0953+414     &   0.2341 &       566 &         44 &       1 &    \mcl{2}{c}{\nodata} & \nodata &    150.1 &     22.1 &       3071 &     27 &          45.2 &        3 &         8.44 &      0.06 &          $-0.40$ \\       
        NGC 3227     &   0.0038 &     14.37 &       0.70 &       1 &      45.0 &        7.2 &       4 &      3.8 &      0.8 &       4112 &    206 &          42.2 &        3 &         7.10 &      0.10 &          $-2.00$ \\       
        NGC 3516$^a$ &   0.0088 &      72.7 &        4.6 &       1 &    \mcl{2}{c}{\nodata} & \nodata &     11.7 &      1.3 &       5384 &    269 &          42.8 &        3 &         7.82 &      0.06 &          $-2.18$ \\       
        NGC 3783     &   0.0097 &      76.3 &       14.1 &      11 &       131 &         20 &       2 &      9.6 &      0.7 &       4486 &     35 &          43.0 &       12 &         7.58 &      0.03 &          $-1.70$ \\       
        NGC 4051     &   0.0023 &     16.30 &       0.57 &       1 &      38.1 &        6.0 &      13 &      2.1 &      0.8 &       1076 &    277 &          41.9 &        3 &         5.68 &      0.28 &          $-0.92$ \\       
        NGC 4151$^a$ &   0.0033 &     46.11 &       0.44 &       1 &      44.1 &    8.3$^c$ &       9 &      6.6 &      1.0 &       6371 &    150 &          42.1 &        3 &         7.72 &      0.07 &          $-2.77$ \\       
          3C 273     &   0.1583 &       409 &         41 &      14 &       675 &        126 &       2 &    146.8 &     10.2 &       3314 &     59 &          45.9 &        3 &         8.50 &      0.03 &             0.28 \\       
        NGC 4593     &   0.0083 &     41.82 &       0.90 &       1 &      54.8 &        8.8 &       4 &      4.0 &      0.8 &       5142 &    572 &          42.6 &        3 &         7.31 &      0.13 &          $-1.84$ \\       
     MCG-6-30-15     &   0.0078 &      19.6 &        4.9 &      15 &    \mcl{2}{c}{\nodata} & \nodata &      5.7 &      1.8 &       1947 &     58 &          41.6 &        3 &         6.63 &      0.14 &          $-2.13$ \\       
        NGC 5548     &   0.0172 &     61.21 &       0.30 &       1 &    \mcl{2}{c}{\nodata} & \nodata &     13.9 &      8.7 &       7256 &   2203 &          43.3 &        3 &         8.15 &      0.38 &          $-2.00$ \\       
         Mrk 817     &   0.0313 &      92.6 &        8.9 &       1 &    \mcl{2}{c}{\nodata} & \nodata &     19.9 &      8.3 &       5348 &    536 &          43.7 &        3 &         8.05 &      0.20 &          $-1.45$ \\       
     PG 1613+658     &   0.1211 &       334 &         40 &       1 &    \mcl{2}{c}{\nodata} & \nodata &     40.1 &     15.1 &       9074 &    103 &          44.8 &        3 &         8.81 &      0.16 &          $-1.19$ \\       
        Z 229-15     &   0.0279 &      20.4 &        5.8 &      16 &    \mcl{2}{c}{\nodata} & \nodata &      3.9 &      0.8 &       3350 &     72 &          42.9 &       17 &         6.93 &      0.09 &          $-1.18$ \\       
         Mrk 509     &   0.0344 &     121.3 &        1.6 &       1 &       297 &         31 &       2 &     79.6 &      5.8 &       3015 &      2 &          44.2 &        3 &         8.15 &      0.03 &          $-1.11$ \\       
        NGC 7469     &   0.0163 &     85.29 &       0.43 &       1 &    \mcl{2}{c}{\nodata} & \nodata &     10.8 &      2.4 &       4369 &      6 &          43.5 &        3 &         7.60 &      0.10 &          $-1.24$ \\\hline 
        NGC 1365$^a$ &   0.0055 &    \mcl{2}{c}{\nodata} & \nodata &      38.1 &        4.8 &       2 & \mcl{2}{c}{\nodata} &       1586 &    465 &          41.9 & 2$^f$,18 &      \mcl{2}{c}{\nodata} &          \nodata \\       
 IRAS 03450+0055     &   0.0315 &     157.4 &        5.9 &       1 &    \mcl{2}{c}{\nodata} & \nodata & \mcl{2}{c}{\nodata} &       3098 &     55 &          43.9 &     1,19 &      \mcl{2}{c}{\nodata} &          \nodata \\       
 IRAS 09149-6206$^d$ &   0.0573 &    \mcl{2}{c}{\nodata} & \nodata &       482 &         49 &       2 & \mcl{2}{c}{\nodata} &       4281 &    121 &          45.0 &    20,21 &         8.06 &      0.25 &          $-0.26$ \\       
        Mrk 1239     &   0.0199 &    \mcl{2}{c}{\nodata} & \nodata &       189 &         30 &       4 & \mcl{2}{c}{\nodata} &        830 &     10 &          44.5 &   22$^f$ &      \mcl{2}{c}{\nodata} &          \nodata \\       
         WPVS 48     &   0.0370 &      70.8 &        4.6 &      23 &    \mcl{2}{c}{\nodata} & \nodata & \mcl{2}{c}{\nodata} &       1890 &     60 &          43.6 &       24 &      \mcl{2}{c}{\nodata} &          \nodata \\       
         Mrk 744     &   0.0091 &      19.9 &        2.2 &       1 &    \mcl{2}{c}{\nodata} & \nodata & \mcl{2}{c}{\nodata} &       5616 &    129 &          41.8 & 1,21$^g$ &      \mcl{2}{c}{\nodata} &          \nodata \\       
  HE~1029$-$1401     &   0.0858 &    \mcl{2}{c}{\nodata} & \nodata &       880 &        133 &       4 & \mcl{2}{c}{\nodata} &       5684 &    284 &          44.6 &       21 &      \mcl{2}{c}{\nodata} &          \nodata \\
          GQ Com     &   0.1650 &       210 &         40 &      25 &    \mcl{2}{c}{\nodata} & \nodata & \mcl{2}{c}{\nodata} &       5036 &    252 &          44.6 &       26 &      \mcl{2}{c}{\nodata} &          \nodata \\       
         Mrk 231     &   0.0422 &    \mcl{2}{c}{\nodata} & \nodata &       393 &         83 &      13 & \mcl{2}{c}{\nodata} &       3130 &    156 &          45.0 & 2$^f$,27 &      \mcl{2}{c}{\nodata} &          \nodata \\       
     ESO 323-G77     &   0.0150 &    \mcl{2}{c}{\nodata} & \nodata &     100.0 &        4.8 &      28 & \mcl{2}{c}{\nodata} &       2635 &    132 &          43.1 &       21 &      \mcl{2}{c}{\nodata} &          \nodata \\       
 IRAS 13349+2438     &   0.1076 &    \mcl{2}{c}{\nodata} & \nodata &      1096 &         71 &      13 & \mcl{2}{c}{\nodata} &       1796 &     90 &          45.0 &       29 &      \mcl{2}{c}{\nodata} &          \nodata \\       
        IC 4329A     &   0.0161 &    \mcl{2}{c}{\nodata} & \nodata &       178 &         10 &       4 & \mcl{2}{c}{\nodata} &       6472 &    324 &          43.2 &       21 &      \mcl{2}{c}{\nodata} &          \nodata \\       
       PGC 50427$^e$ &   0.0235 &      46.7 &        2.2 &      30 &    \mcl{2}{c}{\nodata} & \nodata & \mcl{2}{c}{\nodata} &       3036 &     74 &          43.1 &       30 &         7.34 &      0.04 &          $-1.42$ \\       
         PDS 456     &   0.1840 &    \mcl{2}{c}{\nodata} & \nodata &      1599 &        213 &       2 & \mcl{2}{c}{\nodata} &       3974 &    764 &          46.3 &       31 &      \mcl{2}{c}{\nodata} &          \nodata \\       
       PGC 89171     &   0.0270 &    \mcl{2}{c}{\nodata} & \nodata &       303 &         36 &       4 & \mcl{2}{c}{\nodata} &       2644 &    132 &          43.9 &       21 &      \mcl{2}{c}{\nodata} &          \nodata \\
        NGC 7603     &   0.0288 &    \mcl{2}{c}{\nodata} & \nodata &       332 &         66 &       4 & \mcl{2}{c}{\nodata} &       6350 &    318 &          44.4 &       21 &      \mcl{2}{c}{\nodata} &          \nodata \\ \hline 
\end{tabular}
\end{center}
{\footnotesize 
\begin{itemize}
\item[$^a$] This target is discussed as a changing-look quasar in the literature.
\item[$^b$] Mrk~590 is a changing look AGN displaying strong variability over 
the time of BLR and dust continuum observations \citep{Denney2014}.  We do not 
include it in our statistical analysis.
\item[$^c$] The continuum radius of NGC~4151 reported by 
\cite{Kishimoto2011} is statistically consistent with the recent measurement 
from \cite{Kishimoto2022}.  We prefer the early measurements because it is close 
to the RM measurement.
\item[$^d$] The BLR of IRAS~09149$-$6206 was resolved by GRAVITY and the BH mass 
is derived by \cite{GC2020iras}. The \lopt\ and FWHM of \hb\ are from the BASS 
catalog \citep{Koss2017}.
\item[$^e$] The BH mass of PGC~50427 was measured by RM of the \ha\ line 
\citep{PozoNunez2015}.  We quote the \lopt\ from \cite{Probst2020} observed 
close in time to \ha\ RM.  We measure the FWHM of \hb\ line from the 6dF 
spectrum \citep{Jones2009}.
\item[$^f$] We converted the bolometric luminosity to \lopt\ using Equation~(A.2) of \cite{GC2020cont}.
\item[$^g$] \hb\ FWHM is converted from \ha\ FWHM assuming $\mathrm{FWHM_{H\beta}}/\mathrm{FWHM_{H\alpha}}=1.17$ \citep{Greene2005}.
\end{itemize}
\textbf{Notes.} 
Col.~(1): Target name. 
Col.~(2): Redshift from NASA/IPAC Extragalactic Database (NED). 
Col.~(3): Dust continuum radius based on RM measurement.
Col.~(4): Reference of \Rd\ (RM).
Col.~(5): Dust continuum radius based on OI measurement.
Col.~(6): Reference of \Rd\ (OI).
Col.~(7): BLR radius based on \hb\ time lag.
Col.~(8): \hb\ FWHM.
Col.~(9): AGN optical luminosity at 5100~\AA.
Col.~(10): References of \Rb, FWHM, and \lopt.
Col.~(11): BH mass derived from \Rb\ and FWHM from Columns (7) and (8) assuming the virial factor $f=1$.
Col. (12): Eddington ratio derived from \lopt\ and BH mass from Columns (9) and (11) with the bolometric correction factor 9 \citep{Peterson2004}.
\vspace{1mm}
\\
\textbf{References:} 
(1) \cite{Minezaki2019}, 
(2) \cite{GC2020cont}, 
(3) \cite{Du2019}, 
(4) GRAVITY Collaboration (in preparation),
(5) \cite{Huang2019},
(6) \cite{Ramolla2018},
(7) \cite{Mandal2018},
(8) \cite{Stalin2011},
(9) \cite{Kishimoto2011},
(10) \cite{Du2018},
(11) \cite{Lira2011},
(12) \cite{Bentz2021},
(13) \cite{Kishimoto2009},
(14) \cite{Figaredo2020},
(15) \cite{Lira2015},
(16) \cite{Mandal2021a},
(17) \cite{Barth2011},
(18) \cite{Onori2017}, 
(19) \cite{Rashed2015},
(20) \cite{GC2020iras},
(21) \cite{Koss2017},
(22) \cite{Pan2021},
(23) \cite{PozoNunez2014},
(24) \cite{Probst2020},
(25) \cite{Sitko1993}, 
(26) \cite{Shangguan2018},
(27) \cite{Zheng2002},
(28) \cite{Leftley2021},
(29) \cite{Dong2018},
(30) \cite{PozoNunez2015},
(31) \cite{Nardini2015}.
}
\label{tab:smp}
\end{table*}

\subsection{Dust continuum measurements}
\label{ssec:cont}

We collect type~1 AGNs with dust continuum sizes measured by RM and/or OI in 
Table~\ref{tab:smp}.  The dust continuum sizes based on $K$-band RM observations 
are mainly measured by \cite{Koshida2014}.  
\cite{Minezaki2019} summarized the results of \cite{Koshida2014} in their 
Table~6, using the power index $\alpha=0.1$ to remove the NIR emission from 
the accretion disk (i.e. $f_\nu \propto \nu^\alpha$) to be consistent 
with their primary results.  The assumption of 
the power index may introduce a $\lesssim 10\%$ difference in the time lag, 
which is typically not significant compared to the measurement uncertainty 
\citep{Koshida2014,Minezaki2019}.  We, therefore, adopt the time lags from 
\cite{Minezaki2019} whenever available.  The $K$-band time lag probes 
the dust emission size at shorter wavelength for higher redshift sources, so we 
prefer not to include AGNs much higher than $z \approx 0.2$ (see also 
Section~\ref{ssec:dev}).  \cite{Minezaki2019} also reported $K$-band continuum 
RM measurements for a sample of quasars at $z \approx$~0.1--0.6, most of which 
do not have \hb\ RM measurements.  We only include three quasars from this 
sample, PG~0844+349, PG~0953+414, and PG~1613+658, because they have \hb\ RM 
measurements and are valuable to study the relation of the dust 
continuum and BLR sizes.  $K$-band RM measurements of the other targets, 
e.g. 3C~120 \citep{Ramolla2018} and H~0507+164 \citep{Mandal2018}, are collected 
from individual papers.

We also find 23 AGNs with dust continuum sizes measured by OI, which consists 
of Keck \citep{Kishimoto2009,Kishimoto2011} and recent VLTI/GRAVITY observations 
(\citealt{GC2020cont} and in preparation).  Kishimoto et al. measured the dust 
continuum size by fitting squared visibility amplitudes ($V^2$) with a thin-ring 
model.  They corrected the influence of the accretion disk assuming a point 
source contribution to the visibilities.  \cite{GC2020cont} measured the dust 
continuum size by fitting a Gaussian model to the $V^2$.  
They converted the Gaussian FWHM to the thin-ring radius and corrected 
the radius assuming a 20\% contribution of the coherent flux from the accretion 
disk.  Recent GRAVITY observations measured the dust continuum size of seven 
AGNs with $\lesssim 1$~hour observation time (GRAVITY Collaboration in 
preparation).  We follow the method of \cite{GC2020cont} to measure their 
continuum size.  To estimate the uncertainty, we sum in quadrature 
the statistical uncertainty of the size measurements of individual exposures and 
a 10\% systematic uncertainty (\citealt{GC2020cont} and in preparation).

\subsection{BLR measurements}
\label{ssec:blr}

The BLR size can be probed by different broad emission lines.  
The \hb\ line has been the most extensively used in RM campaigns of low-$z$ AGNs 
\citep[e.g.][]{Bentz2015}.  Meanwhile, GRAVITY spectroastrometric observations 
probe the BLR with \paa\ and \brg\ \citep{GC2018,GC2020iras,GC2021ngc3783}.  
Different broad lines of a BLR may show a different size due to 
the photoionization conditions and optical depth \citep{Korista2004}.  
In this work, we only study the relation of the dust continuum radius and \hb\ 
BLR radius from the RM time lag for simplicity.  Meanwhile, future GRAVITY 
observations measuring the BLR and dust continuum simultaneously in $K$-band 
will be powerful to investigate their relation (see Section~\ref{sec:fut}).  
We collect the \hb\ BLR radii of most targets, together with their optical 
luminosities at 5100~\AA\ and \hb\ line FWHMs, from Table~1 of 
\cite{Du2019}.\footnote{Similar data are also collected in \cite{DallaBonta2020}.}  
\cite{Du2019} averaged the quantities if there is more than one measurement.  
We only find a few additional AGNs from the other references (see 
Table~\ref{tab:smp}).  We adjust the \lopt\ to our cosmology when necessary 
unless the distances of some nearby AGNs are explicitly specified in 
the references.

\begin{figure*}
\centering
\includegraphics[height=0.3\textheight]{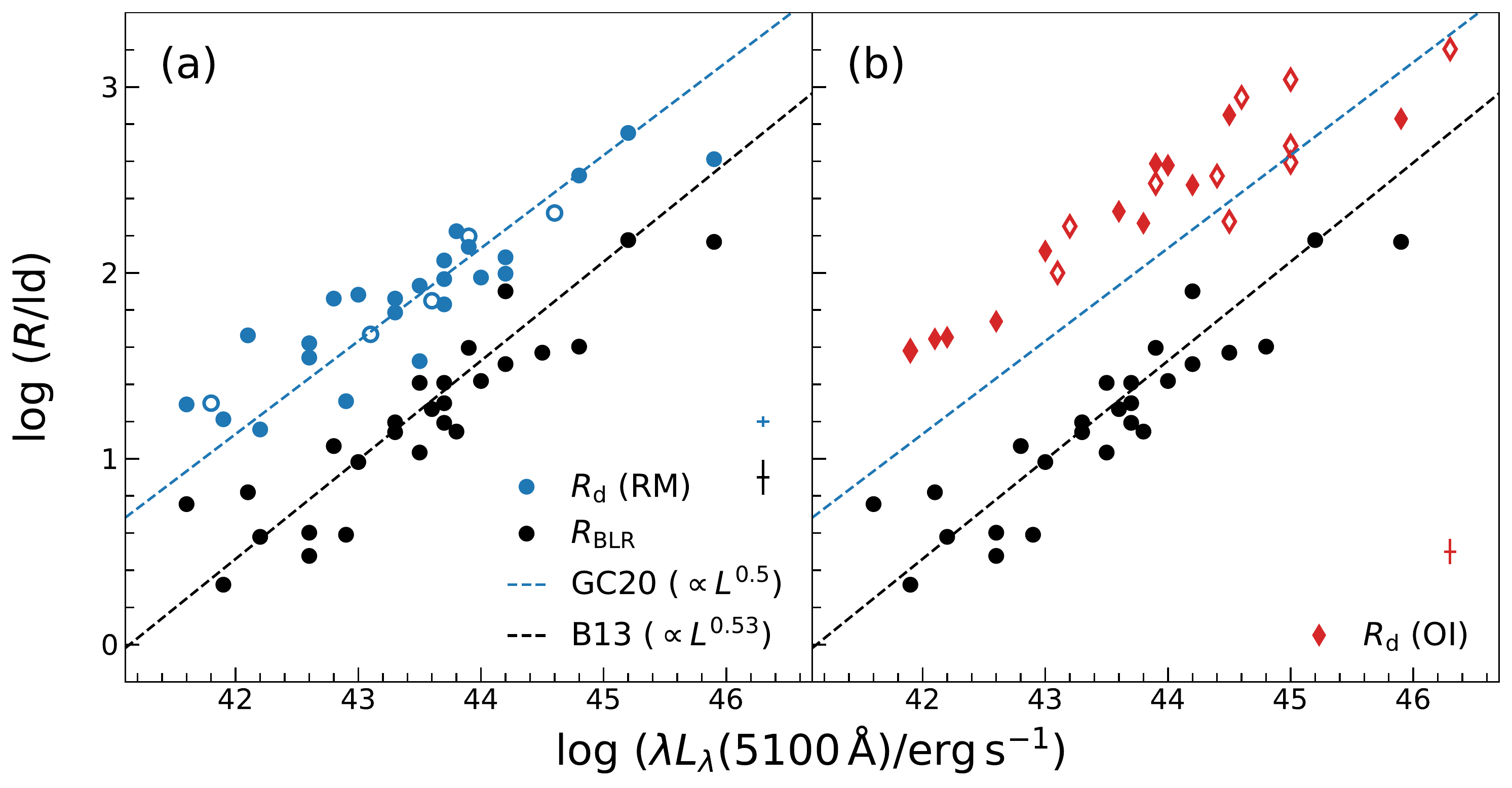}
\caption{Size--luminosity relations of the dust continuum measured by 
(a) RM (blue circles) and (b) OI (red diamonds) are systematically above 
that of the BLR (black circles; same in both panels).  
The open-colored circles indicate sources without BLR measurements.  
The dashed lines in both panels are based on RM-measured \rl\ 
relations of the dust continuum \citep[blue;][]{GC2020cont} and 
the BLR \citep[black;][]{Bentz2013}.  The typical uncertainties are shown on 
the lower right of each panel.} 
\label{fig:sample}
\end{figure*}

\subsection{AGN variability}
\label{ssec:var}

We emphasize that the BLR and dust continuum sizes are usually measured 
in different epochs.  While the BLR RM campaigns of many targets conducted 
from 2000--2010 coincide with dust continuum RM campaigns 
\citep{Koshida2014,Minezaki2019}, the time difference between BLR and dust 
measurements for some targets can be over a decade.  For example, we adopt 
the BLR RM of NGC~3783 measured in 2020 \citep{Bentz2021}, while its dust 
continuum RM was measured around 2009 \citep{Lira2011}.  
We find that the early BLR RM measurements 
of NGC~3783 \citep{Onken2002} yielded a BLR radius and \lopt\ quite close to 
the recent measurements.  We, therefore, prefer to adopt the new measurements 
for simplicity.  The asynchronous effect will contribute to the scatter of the 
size correlation (see discussion in Section~\ref{sec:sc}).

Targets with substantial variability may show a large scatter in the \Rb--\Rd\ 
relation.  We identify so-called `changing-look' AGNs in our sample from the 
literature and discuss them in the following.  In the end, we are convinced 
that only Mrk~590 is not suitable to be included in our analysis.  
Mrk~590 transformed from a classical type~1 AGN to type 1.9--2, as its continuum 
luminosity decreased by 100 over the past four decades \citep{Denney2014}.  
The lack of evidence of intrinsic absorption indicates that changes in continuum 
and emission lines are due to the decline of the BH accretion rate instead of 
obscuration along the line of sight.  The BLR of Mrk~590 was monitored through 
the \hb\ line from 1990--1996 
\citep{Peterson1998} and recently, with \ha\ in 2018 \citep{Mandal2021b}.  
The BLR size of Mrk~590, around 25~ld, does not change significantly at 
the bright and faint states.  The dust continuum RM campaign was 
conducted from 2003 to 2007 \citep{Koshida2014} during the rapid decline of the AGN 
luminosity in Mrk~590.  \cite{Kokubo2020} found that the dust continuum size of 
Mrk~590 is quite small, only $\sim 33$~ld, reflecting the rapid replenishment 
of the dust in the innermost region of the dusty interstellar medium.  
We exclude Mrk~590 in our statistics  to avoid the complicated physics of this 
target.  There are other changing-look AGNs in our sample, e.g. Mrk~6 and 
NGC~4151 (flagged in Table~\ref{tab:smp}).  They change either from type~1 to 
type~2 or vice versa (see \citealt{Marin2019,Senarath2021} and references 
therein).  However, we find the BLR and continuum were measured when they stayed 
the same type, so we include them in our analysis.  Moreover, some 
changing-look AGNs flagged in our sample are likely caused by temporary 
changes in the line-of-sight obscuration \citep[e.g.][]{Goodrich1989,Shapovalova2019}.  
This mechanism does not relate to any intrinsic change in the AGN properties, 
so it will not affect the \Rb--\Rd\ relation that we are interested in for this 
work.  Nevertheless, our results stay the same if we exclude all  
the changing-look AGNs.  \cite{Clavel1989} published their $K$-band RM 
measurement of Fairall~9 while its UV continuum flux was dropping by a factor of $\sim 30$.  However, an \hb\ RM measurement before it changed to 
the faint state does not exist.  The extreme variability, from type~1 to almost type~2 
\citep{Kollatschny1985,Lub1992}, prevents a simple choice of \lopt\ and \hb\ 
FWHM for our analysis.  Therefore, we chose to exclude Fairall 9 in this work.

\section{Scaling relations of BLR and dust continuum size}
\label{sec:sc}

Figure~\ref{fig:sample} displays our sample.  The dust \rl\ relations 
measured by both RM and OI are systematically above the BLR \rl\ relation.  
Moreover, the continuum size measured by OI is above that of RM.  
The \rl\ relations of dust RM and OI measurements have been discussed in many 
previous works
\citep[e.g.][]{Suganuma2006,Kishimoto2009,Kishimoto2011,Koshida2014,Minezaki2019,GC2020cont}.  
One common explanation is that the RM-measured time lag is 
weighted by the amplitude of flux variations, 
which is expected to originate most strongly from the inner boundary of the hot 
dust; on the other hand, OI-measured sizes are mainly flux-weighted and, 
therefore, are elevated by contributions of lower temperature dust at larger 
radii \citep{Kishimoto2009,Kishimoto2011}.  Another slightly different 
explanation assumes that the dust structure is a `bowl shape' 
\citep{Kawaguchi2010}: we mainly observe the foreground side 
at a low inclination angle for type~1 AGNs, and dust at larger radii is also 
closer to the observer, so the projected size increases more significantly than 
their RM time lag towards larger radii \citep{PozoNunez2014,Figaredo2020}.  
A more detailed discussion of BLR and dust structure models is beyond 
the scope of this paper.  Throughout the paper,  we use OI and RM to refer to 
the dust continuum 
measurements from these two different methods unless otherwise clarified.

Previous works studying both RM and OI observations have hinted that the slope of 
the dust continuum \rl\ relation is shallower than $\propto L^{0.5}$. 
The slopes in our current RM and OI samples are both about 0.4, 
consistent with previous works.  We leave a more detailed discussion of the \rl\ 
relation in a separate paper (GRAVITY Collaboration et al. in preparation). 
In summary, previous RM studies provide various explanations of the shallower 
slope relating to the dust structure, dust response to the accretion disk 
emission, and its putative effect on the observed optical luminosity 
\citep[e.g.][]{Minezaki2019,Figaredo2020}.  Based on the OI measurements, 
\cite{GC2020cont} suspect the continuum emission of the accretion disk may 
bias the OI size measurement for the most luminous sources.

Likewise, recent RM BLR measurements \citep{Du2015,Grier2017b} also find targets 
with \Rb\ significantly lower than the canonical BLR \rl\ relation 
\citep[e.g.][]{Bentz2013}.  Some works find that the deviation closely 
correlates with the accretion rate of the BH 
\citep{Du2015,Du2019,MartinezAldama2019,DallaBonta2020}, while 
the physical driver remains unclear in some other works 
\citep{Grier2017b,FonsecaAlvarez2020}.   
The physical explanation of the deviation of BLR and dust continuum \rl\ 
relations is beyond the scope of the current paper.  
As discussed in the following sections, we find tight relations between 
\Rb\ and \Rd, which are close to linear, reflecting a simple link between 
the two physical structures.

\begin{figure*}
\centering
\includegraphics[height=0.3\textheight]{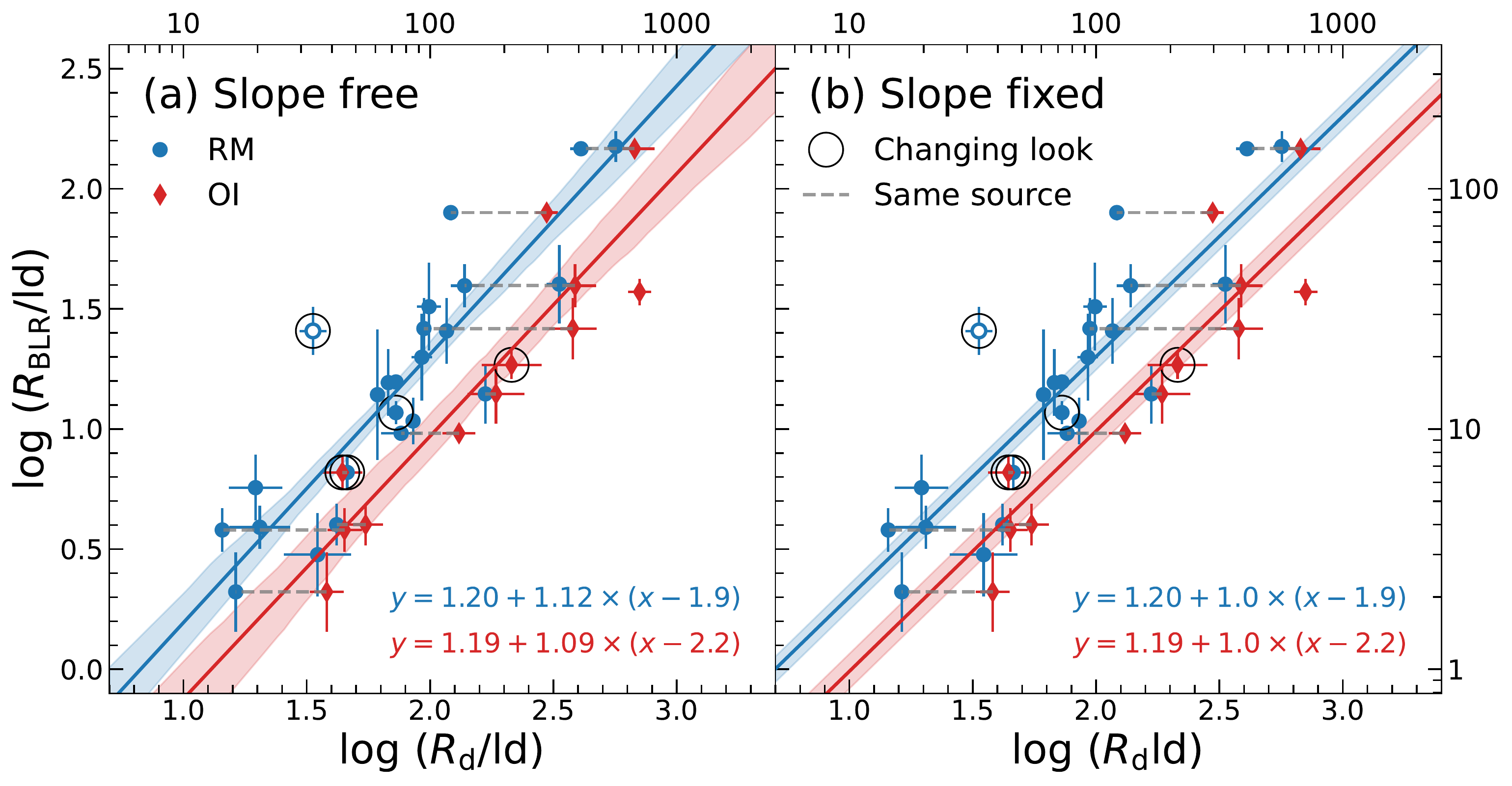}
\caption{Tight relations between the radii of the BLR and the dust continuum.  
The blue circles are AGNs measured by RM, and the red diamonds are the targets 
measured by OI.  The gray dashed lines connect the same sources with 
both RM and OI measured \Rd.  The best-fit \Rb--\Rd\ relations of RM and OI 
datasets (formula in the lower right corner of each panel) are shown as the blue 
and red lines, respectively.  The black circles enclose the known changing-look 
AGNs.  We only exclude Mrk~590, denoted as the empty blue circle, in our 
analysis.  The best-fit results with the slope free (a) or fixed to unity (b) 
are almost the same, so we mainly adopt the results of (b) in the discussion.
The linear scale is indicated on the top and right axes.}
\label{fig:sc}
\end{figure*}

\subsection{Statistical analysis}
\label{ssec:stat}

We find strong correlations between BLR and dust continuum sizes, as shown in 
Figure~\ref{fig:sc}.  We fit the relations for RM- and OI-measured \Rd\ 
separately as they show systematic differences.  We fit the data with a 
power-law relation,
\begin{equation}\label{eq:mod}
\log\,(\Rb/\mathrm{ld}) = \alpha + \beta \log\,(\Rd/R_\mathrm{d,0}),
\end{equation}
where $\alpha$ and $\beta$ are the intercept and the slope, and
$R_\mathrm{d,0}$ is a pivot point fixed close to the median of \Rd\ of the data 
to reduce the degeneracy of $\alpha$ and $\beta$.  We adopt the Bayesian Markov 
Chain Monte Carlo (MCMC) approach to fit the data.  The likelihood is,
\begin{equation}\label{eq:lnl}
\ln\,\mathcal{L} = -\frac{1}{2} \Sigma_i \left(\ln (2\pi \sigma_i^2) + \frac{(y_i - m_i)^2}{\sigma_i^2} \right),
\end{equation}
where $y_i$ is $\log\,\Rb$ data, $m_i$ is the model value based on 
$\log\,\Rd$ and Equation~(\ref{eq:mod}), and 
$\sigma_i^2 = (\beta\, \sigma_{x, i})^2 + \sigma_{y, i}^2 + \epsilon^2$ 
includes the measurement uncertainties of the dust continuum ($\sigma_{x,i}$) and 
BLR ($\sigma_{y,i}$) radius as well as the intrinsic scatter ($\epsilon$).  
We adopt uniform priors of the parameters that are wide enough and sample 
the posterior with the widely used Python package of MCMC, \texttt{emcee} 
\citep{dfm2013}.  We used 32 walkers and 5000 steps, with the first 500 steps 
discarded as burn-in steps.  The fitting is well converged.  

\begin{table}
\caption{Best-fit parameters of Equation~(\ref{eq:mod})}
\renewcommand{\arraystretch}{1.5}
\small
\begin{center}
\begin{tabular}{r c c c c}
\hline\hline
  Relation &               $\alpha$ &                $\beta$ &             $\epsilon$ & $\log\,(R_\mathrm{d,0}/\mathrm{ld})$ \\ \hline
RM (free)  & $1.20^{+0.05}_{-0.05}$ & $1.12^{+0.13}_{-0.13}$ & $0.21^{+0.05}_{-0.04}$ &                                  1.9 \\ 
OI (free)  & $1.19^{+0.08}_{-0.08}$ & $1.10^{+0.18}_{-0.18}$ & $0.25^{+0.08}_{-0.06}$ &                                  2.2 \\ \hline
RM (fixed) & $1.20^{+0.05}_{-0.05}$ &                      1 & $0.21^{+0.04}_{-0.04}$ &                                  1.9 \\ 
OI (fixed) & $1.19^{+0.08}_{-0.08}$ &                      1 & $0.24^{+0.07}_{-0.05}$ &                                  2.2 \\ \hline
\end{tabular}
\end{center}
{\textbf{Notes.} 
We fit the \Rb--\Rd\ relation for AGNs with \Rd\ measured by RM and OI, 
respectively.  $\alpha$, $\beta$, and $\epsilon$ are the intercept, slope, and 
intrinsic scatter of a linear relation. $R_\mathrm{d,0}$ is the pivot point 
fixed in the fitting.  The first two rows provides the best-fit parameters with 
$\beta$ free, while the last two rows are results with $\beta$ fixed to unity. 
OI-measured \Rd\ is about 0.3~dex larger than that measured by the RM. 
}
\label{tab:fit}
\end{table}

The best-fit relations are very close to linear (throughout the paper, 
`linear' refers $\beta=1$) when we allow the slope to be free, as shown in 
Figure~\ref{fig:sc}a.  Therefore, we also fit the data with the slope fixed to 
unity (Figure~\ref{fig:sc}b).  The best-fit results are listed in Table~\ref{tab:fit}.  
For simplicity, we will take the $\beta$-fixed fitting results in the following 
discussion and derive the BLR radius and BH mass with \Rd\ in Section~\ref{sec:mbh}.  
Given current uncertainties, whether we adopt the relations with $\beta$ free or 
fixed does not affect these results.  

The dust continuum size measured from RM is about 0.7~dex (five times) 
larger than the BLR size, which is consistent with previous works 
\citep[e.g.][]{Koshida2014,Kokubo2020}.  The dust continuum size measured by OI 
is about 0.3~dex (two times) larger than that measured by RM, again 
consistent with previous studies \citep{Kishimoto2011,Koshida2014,GC2020cont}.  
We do not find the slope of the \Rb--\Rd\ relation to significantly deviate 
from unity.  However, for the relation to be linear, any departure from 
$\propto L^{0.5}$ for the \rl\ relations of the dust continuum and the BLR must 
be similar.  Our sample shows a more significant deviation in \Rd--\lopt\ than 
that in \Rb--\lopt.  This difference may contribute to the scatter of 
the \Rb--\Rd\ relation, which we will discuss in Section~\ref{ssec:dev}.

\subsection{Intrinsic scatter}
\label{ssec:sig}

The intrinsic scatter of the best-fit relations is about 0.25~dex for both RM 
and OI relations.  The physical difference between the BLR and dust structure 
for individual targets may contribute to the scatter.  Such a variation was 
observed in the mid-IR, where a large scatter of the \rl\ relation was observed 
\citep{Burtscher2013}.  However, besides the individual BLR and dust structure 
difference, for the \rl\ relation, the uncertainty of the bolometric luminosity 
of an AGN is another primary source of the scatter.  Studying the BLR and dust 
structure with the \Rb--\Rd\ relation allows us to avoid the uncertain 
bolometric luminosity.  However, AGN variability still likely introduces 
considerable intrinsic scatter because the BLR and dust continuum sizes are not 
measured in a state where they reflect the same AGN bolometric 
luminosity.  

As described in Section~\ref{ssec:var}, most of the dust continuum RM 
measurements are made between 2001 and 2008, while the OI measurements are 
around 2009--2010 (Keck interferometry) and 2018--2022 (GRAVITY).  
The BLR measurements are conducted from the 1980s until recent years. 
NGC~5548 is one of the best targets to investigate the variability: 
\cite{Du2019} collected 18 epochs of its BLR RM measurements from 1989 to 2015,  
and the BLR radius (time lag) varies from 4.2~ld to 26.5~ld with a standard 
deviation of 0.24~dex.  Mrk~335 and Mrk~817 also have four epochs of BLR 
measurements across $>10$~years, and their BLR radii changes are 
$\gtrsim 0.3$~dex.  For the dust continuum, \cite{Koshida2014} reported 6~epochs 
of dust continuum RM measurements for NGC~5548 from 2001 to 2007; 
the RMS of the dust continuum radius is about 0.1~dex. Other AGNs, NGC~3227, 
NGC~4051, and NGC~4151, with $\geq 4$ epochs of dust RM measurements, also show 
similar $\sim 0.1$~dex RMS variation.  NGC~3783 was observed from 1974--1990 
\citep{Glass1992} and later in 2006--2009 \citep{Lira2011}; 
the measured $K$-band time lags are consistent within their uncertainties.  
It is not surprising that the dust continuum size shows less variability: 
the dust re-radiation effectively averages the variability of the central engine 
on a longer time scale, while the RM technique measures the averaged size over 
the monitoring period.

Since the BLR size variation of the AGNs with multiple measurements always reach 
$\gtrsim 0.2$~dex, we conclude that the observed 0.25~dex intrinsic scatter of 
the \Rb--\Rd\ relations can be explained by the time variation of the BLR and dust 
continuum sizes, while the physical difference between the BLR and dust structures of 
individual targets also plays a role.  
Future simultaneous measurements of BLR and dust continuum sizes have the potential to reveal a tighter \Rb--\Rd\ relation, while the related caveats are 
discussed in Section~\ref{ssec:cav}.

\subsection{Higher-order correlations}
\label{ssec:dev}

\begin{figure*}
\centering
\includegraphics[height=0.3\textheight]{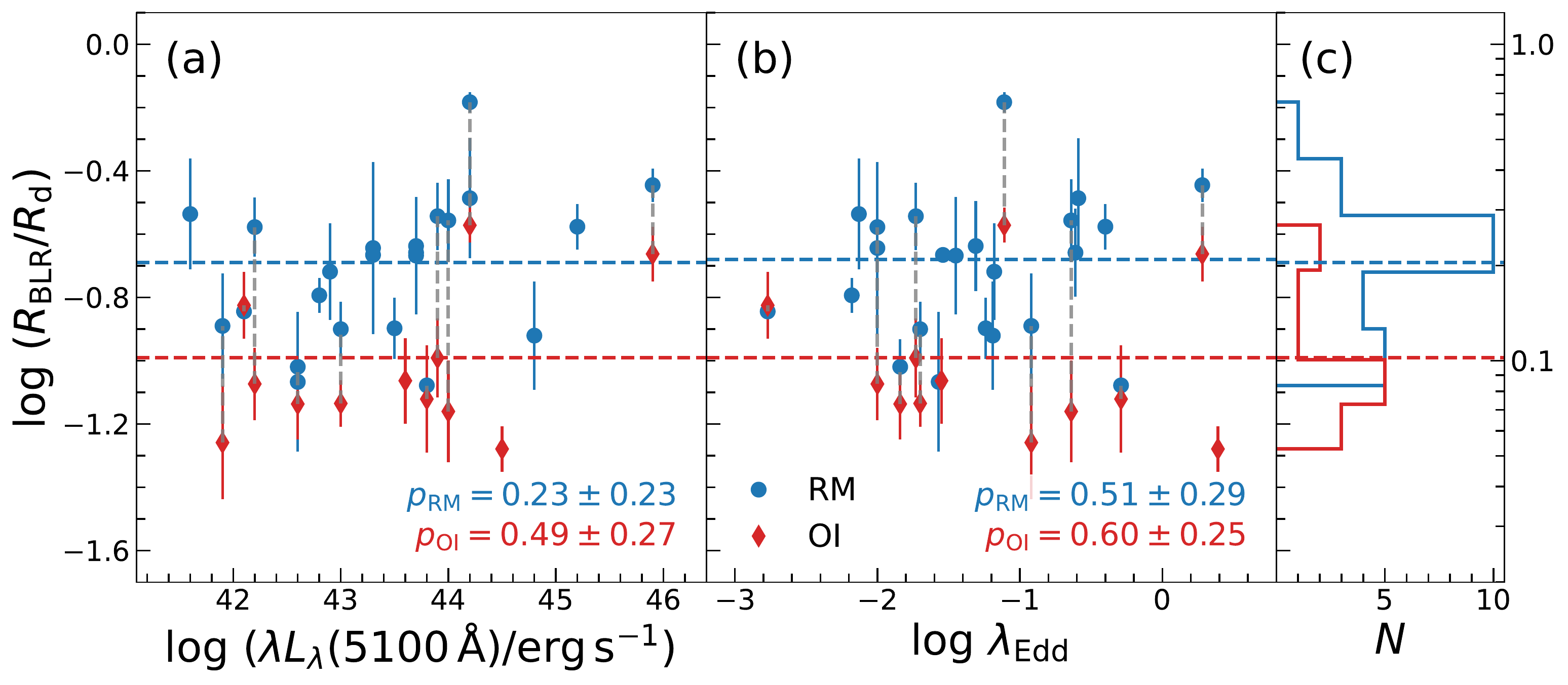}
\caption{Ratio of BLR and dust continuum sizes as a function of (a) the AGN 
luminosity and (b) the Eddington ratio.  The $p$-values of Spearman's rank 
correlation coefficient are reported in the lower right corner of each panel for 
RM- and OI-measured samples.  The notations are the same as in 
Figure~\ref{fig:sc}.  The error bars are the quadrature sum of the uncertainties 
of \Rb\ and \Rd.  Panel (c) displays the histograms of the ratios.  The dashed 
lines indicate the averaged $\Rb/\Rd$ based on the fitting of the \Rb--\Rd\ 
relations with $\beta$ fixed to unity (see also Equation~(\ref{eq:rst})).  
$\Rb/\Rd$ does not show a statistically significant correlation with the \lopt\ 
nor the Eddington ratio.  The linear scale of $\Rb/\Rd$ is indicated on 
the right.}
\label{fig:rat}
\end{figure*}

We investigate whether the scatter of the \Rb--\Rd\ relation correlates 
with the physical properties of the AGN.  Since the \Rb--\Rd\ relations are 
remarkably close to linear (Section~\ref{ssec:stat}), we study the dependence of 
the ratio, $\Rb/\Rd$,  on the other physical parameters of the AGN.  
We calculate Spearman's rank correlation coefficients to test the significance 
of the correlations.  We perturb $\Rb/\Rd$ 500 times with the measurement 
uncertainties of \Rb\ and \Rd\ to calculate the $p$-value distribution.  
The resulting $p$-values do not support any significant correlations (i.e. 
$p>0.05$).  We find that the conclusions do not change when we study 
the deviation of the \Rb--\Rd\ relations from the best-fit results with $\beta$ 
free.

We first investigate $\Rb/\Rd$ against the AGN luminosity and 
the Eddington ratio (Figure~\ref{fig:rat}).  Although the $p$-values do not 
support significant correlations, we notice that, at 
$\lopt>10^{45}$~\ergs, the three data points of PG~0953+414 and 3C~273 are all 
above the averaged values of $\Rb/\Rd$.  This trend may drive the slope to 
values slightly larger than 1 when we fit the \Rb--\Rd\ relation with the slope 
free.  Unfortunately, our current sample has too few luminous AGNs to confirm 
such a trend. Previous BLR RM studies have found that the Eddington ratio may 
drive the deviation of the \rl\ relation such that highly accreting AGNs display 
shorter time lags \citep[e.g.][]{Du2015}.  We do not find a dependence of 
$\Rb/\Rd$ on the Eddington ratio.  Nevertheless, our targets do not show 
significant deviation from the \cite{Bentz2013} relation either (Figure~\ref{fig:sample}).
It is worth noting that Mrk~509, which displays the highest deviation of 
$\Rb/\Rd$ has only $\lopt \approx 10^{44.2}$~\ergs\ and intermediate Eddington 
ratio.  More observations are needed to understand the details of 
the \Rb--\Rd\ relation.

\begin{figure*}
\centering
\includegraphics[width=0.9\textwidth]{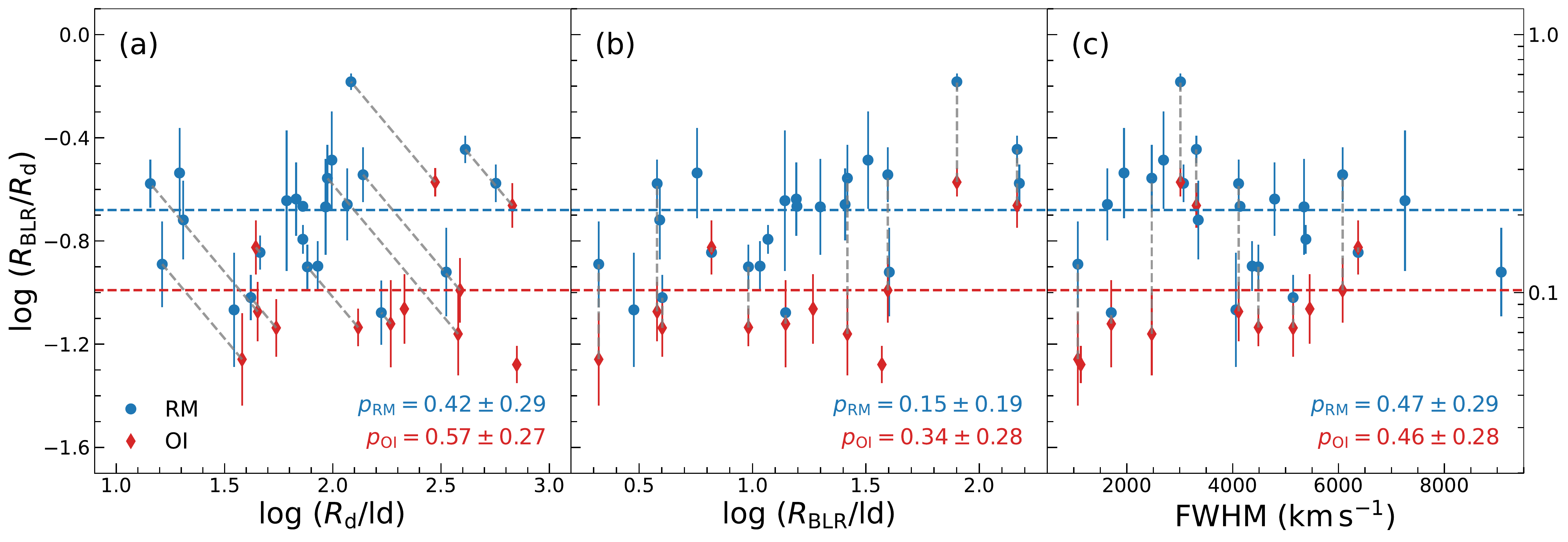}
\caption{Ratio of \Rb\ and \Rd\ as a function of    
(a) the continuum radius, 
(b) the BLR radius, and 
(c) the FWHM of \hb\ line.  
We do not find any statistically significant correlation.
However, all of the AGNs with $\Rb \gtrsim 50$~ld show $\Rb/\Rd$ above 
the averaged values.  We discuss this tentative trend in the text.
The notations are the same as in Figure~\ref{fig:rat}.  
The linear scale of $\Rb/\Rd$ is indicated on the right.}
\label{fig:dev}
\end{figure*}

We further investigate the dependence of $\Rb/\Rd$ on \Rd, \Rb, and 
the FWHM of \hb\ (Figure~\ref{fig:dev}).  Again, Spearman's rank correlation 
coefficients do not support a significant correlation with any of the three 
parameters.  We only find a tentative trend that all three 
targets, PG~0953+414, 3C~373, and Mrk~509, with $\Rb \gtrsim 50$~ld in 
Figure~\ref{fig:dev}b show $\Rb/\Rd$ above the averaged values.  This trend is 
similar to what is discussed above for $\lopt>10^{45}$~\ergs\ AGNs, although we 
caution that the intrinsic scatter of the \Rb--\Rd\ relation will naturally lead 
to the correlation between $\Rb/\Rd$ and $\Rb$.

Moreover, in our sample, AGNs with the largest radii are at $z\gtrsim 0.1$. 
Their $K$-band measurements probe continuum emission at a slightly shorter 
wavelength ($\lesssim 2\,\mu$m) than the rest of the sample.  One may simply 
expect the continuum size to be smaller at shorter wavelengths because 
of a larger contribution from higher temperature dust
that is closer to the central engine \citep[e.g.,][]{Oknyansky2015}. 
Indeed, a sharp decrease in the time lag towards shorter wavelength was observed 
for NGC~4151 \citep{Oknyanskij1999} and GQ Com \citep{Sitko1993}.  
However, it is more common that the time lags of dust emission only decrease 
moderately towards shorter wavelength \citep{Oknyansky2015}.  
A bi-conical dust distribution can explain this because dust with different 
temperatures is located on similar isodelay surfaces.  Therefore, we suggest 
that the redshift effect is not likely to cause the observed deviation. However,  
it is hard to draw a firm conclusion with the limited number of measurements.  
More observations of AGNs with high luminosity and/or at high redshift 
would be essential to investigate this problem further.

\section{Estimating BH masses with dust continuum sizes}
\label{sec:mbh}

\subsection{BLR radius based on dust continuum measurements}
\label{ssec:sc}

We can now use the measured dust continuum radius to estimate the BLR radius 
for AGNs lacking BLR measurements.  We adopt the best-fit parameters with 
$\beta=1$ in Table~\ref{tab:fit} so that Equation~(\ref{eq:mod}) can be written as 
\begin{align}\label{eq:rst}
\log\,\Rb =
  \begin{cases}
    \log\,\Rd - 0.70       & \text{(RM)}, \\
    \log\,\Rd - 1.01       & \text{(OI)}.
  \end{cases}
\end{align}
We adopt 0.25~dex as the uncertainty of \Rb\ based on the intrinsic 
scatter of the \Rb--\Rd\ relation.  We expect the uncertainty of the \Rb--\Rd\ 
relations to be reduced in the future with more observations of the BLR and 
continuum close in time and of more luminous AGNs with large BLR sizes 
(however, see Section~\ref{ssec:cav}).

\begin{figure}
\centering
\includegraphics[width=0.45\textwidth]{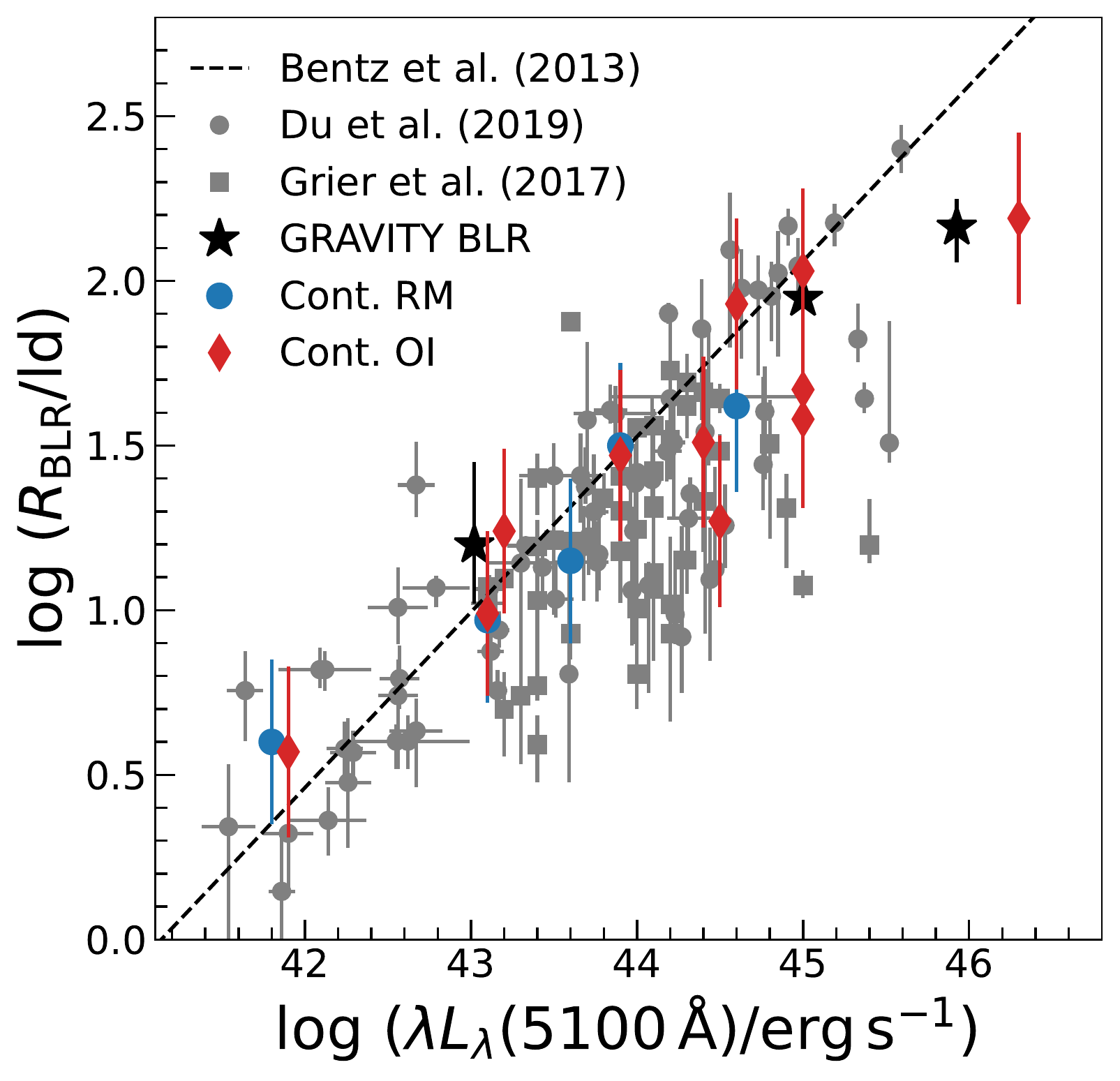}
\caption{BLR \rl\ relation of AGNs measured by the continuum-based 
method.  The blue circles and red diamonds are dust radii measured by RM and OI, 
respectively.  The black stars are three AGNs whose BLR kinematics are resolved 
by GRAVITY \citep{GC2018,GC2020iras,GC2021ngc3783}.  
The gray circles and squares are BLR RM measured AGNs from \cite{Du2019} and 
\cite{Grier2017b} respectively.  
The dashed line is the best-fit \rl\ relation from \cite{Bentz2013}. 
The continuum-based method derives the BLR radii following a similar 
trend of the \rl\ relation based on RM and GRAVITY BLR measurements.} 
\label{fig:rl}
\end{figure}

In Table~\ref{tab:der}, we report the \Rb\ derived from Equation~(\ref{eq:rst}) 
for AGNs in the lower part of Table~\ref{tab:smp}.  We plot these 
targets in Figure~\ref{fig:rl} with their \Rb\ against \lopt.  
The continuum-based \Rb\ follows the \rl\ 
relation of the direct BLR measurements.  Similar to \cite{Du2019} and 
\cite{Grier2017b}'s results, we find AGNs more likely below the \cite{Bentz2013} 
relation, especially for the luminous ($\lopt>10^{44}\,\ergs$) objects.  Although 
we are still limited by the small number of objects, the distribution of our 
targets closely resembles the RM measured sample.  Four targets, Mrk~1239, 
Mrk~231, IRAS~09149$-$6206, and PDS~456, show the most significant deviation 
from the \cite{Bentz2013} relation.  As discussed in Section~\ref{ssec:mass}, 
all of them, except for IRAS~09149$-$6206, are at or above or above their Eddington 
luminosity.  The Eddington ratio of IRAS~09149$-$6206 (about 0.4) is also 
high among the typical AGNs (e.g. the rest of the sample).  Thus the large 
deviations from the \rl\ relation could be related to their high accretion 
rates \citep{Du2019}.

\begin{table}
\caption{BH mass and Eddington ratio derived with dust continuum size}
\renewcommand{\arraystretch}{1.2}
\small
\begin{center}
\begin{tabular}{l l@{$\pm$}r c r}
\hline\hline
\multicolumn{1}{c}{Name} & \mcl{2}{c}{$\log\,\Rb$} & $\log\,\mbh$ & $\log\,\rEdd$   \\ 
                         & \mcl{2}{c}{(ld)}        & (\msun)      &                 \\ 
\multicolumn{1}{c}{(1)}  & \mcl{2}{c}{(2)}         & (3)          & \mcl{1}{c}{(4)} \\\hline
       NGC 1365          & 0.57       & 0.26       & 6.26         & $-1.51$         \\ 
IRAS 03450+0055          & 1.50       & 0.25       & 7.77         & $-1.02$         \\ 
IRAS 09149$-$6206        & 1.67       & 0.25       & 8.23         & $-0.37$         \\
       Mrk 1239          & 1.27       & 0.26       & 6.40         &   0.96          \\ 
        WPVS 48          & 1.15       & 0.25       & 6.99         & $-0.54$         \\ 
 HE 1029$-$1401          & 1.93       & 0.26       & 8.73         & $-1.28$         \\
        Mrk 744          & 0.60       & 0.25       & 7.39         & $-2.73$         \\ 
         GQ Com          & 1.62       & 0.26       & 8.32         & $-0.86$         \\ 
        Mrk 231          & 1.58       & 0.27       & 7.87         & $-0.01$         \\ 
    ESO 323-G77          & 0.99       & 0.25       & 7.12         & $-1.17$         \\ 
IRAS 13349+2438          & 2.03       & 0.25       & 7.83         &  $0.02$         \\ 
       IC 4329A          & 1.24       & 0.25       & 8.17         & $-2.10$         \\ 
      PGC 50427          & 0.97       & 0.25       & 7.22         & $-1.27$         \\ 
        PDS 456          & 2.19       & 0.26       & 8.68         &   0.47          \\ 
      PGC 89171          & 1.47       & 0.26       & 7.61         & $-0.85$         \\ 
       NGC 7603          & 1.51       & 0.26       & 8.41         & $-1.15$         \\ \hline
\end{tabular}
\end{center}
{\textbf{Notes.} 
Col.~(1): Target name. 
Col.~(2): BLR radius derived from \Rd\ with Equation~(\ref{eq:rst}).  However, 
we find the difference between the results using the best-fit relations with $\beta$ 
fixed and free is much smaller than the uncertainties.
Col.~(3): BH mass derived from \Rb\ and \hb\ FWHM assuming the virial factor $f=1$.
Col.~(4): Eddington ratio derived from \lopt\ (Col. (9) of Table~\ref{tab:smp}) and BH mass from Col.~(3).
}
\label{tab:der}
\end{table}

\subsection{BH mass}
\label{ssec:mass}

Since our continuum-based \Rb\ is fully consistent with RM-measured \Rb, one can 
simply adopt Equation~(\ref{eq:vir}) and the virial factor previously calibrated 
for the RM method to derive the BH mass.  We adopt $f=1$ to be consistent with 
\cite{Du2019}.  The FWHM of the \hb\ line can be obtained from a single-epoch 
spectrum (Table~\ref{tab:smp}).  We further calculate the Eddington ratios of 
these targets with the bolometric luminosities scaled from \lopt\ by a bolometric 
correction factor of 9 \citep{Peterson2004}.  Four AGNs, Mrk~1239, Mrk~231, 
IRAS~13349+2438, and PDS~456, are close to or above the Eddington accretion.  Their 
optical spectra (see their references in Col. (10) of Table~\ref{tab:smp}) 
commonly show features of narrow widths of the broad \hb\ lines, weak or no 
\OIIIc\ lines, and strong \feii\ features.  These features strongly indicate 
that these targets have high accretion rates 
(\citealt{Boroson1992,Shen2014} and references therein), consistent 
with our Eddington ratios.  The BH masses of IRAS~09149$-$6206 and 
PGC~50427 are $10^{8.06}\,\msun$ and $10^{7.34}\,\msun$, based on 
the spectroastrometry of the broad \brg\ line \cite{GC2020iras} and the RM of \ha\ 
line \citep{PozoNunez2015}, respectively.  Our derived BH masses by 
the continuum-based method are very close ($<0.2$~dex) to those from the direct 
BLR measurements.

The primary uncertainty of the RM method to measure the BH mass comes from 
the virial factor.  The calibration of $f$ typically shows $\sim 0.4$~dex 
intrinsic scatter \citep{Woo2010,Ho2014}, which consists of the variation of 
the BLR structure of individual targets and the intrinsic scatter of 
the \mbh--$\sigma_*$ relation \citep{Gebhardt2000,Ferrarese2000}.  
The sample selection and regression method may introduce a factor of 2 
systematic difference of the virial factor \citep{Graham2011,Park2012}.  
The virial factor may differ by a factor of 2 depending on the bulge 
type (classical bulge and pseudo bulge; \citealt{Ho2014}), which might be 
correlated to the systematics of the sample selection.  For individual AGNs, 
one way to measure their virial factors is via 
dynamically modeling the velocity-resolved RM data 
\citep{Pancoast2014a,Pancoast2014b}.  Based on dynamical modeling 
results, \cite{Williams2018} find 0.2--0.5~dex uncertainties on 
the predictive distribution of the virial factor corresponding to different 
definitions of the line width.  One should be cautious that the small scatter 
(e.g. 0.2~dex) may be due to the narrow range of parameter space spanned by 
their small sample.  Altogether, to derive the BH mass, the uncertainty of 
the virial factor is likely $\gtrsim 0.3$~dex.  Since the continuum-based method 
shares the same uncertainty of the virial factor as the RM method, 
the \Rb--\Rd\ relation provides a promising method to measure the BH mass close, 
if not equivalent, to the accuracy of the RM method.  

The single-epoch method is generally believed to be much more uncertain than RM due to the intrinsic scatter and the systematic bias of 
the \rl\ relation.  The systematic deviation of the measured \Rb\ from the 
canonical \rl\ relation is discussed in Section~\ref{sec:sc} and 
Section~\ref{ssec:sc}.  \cite{DallaBonta2020} recently provided 
the state-of-the-art calibration of the single-epoch method with the RM database 
\citep{Bentz2015} and SDSS RM \citep{Grier2017b} samples.  
Their calibration empirically includes the secondary dependence of the Eddington 
ratio which results in an intrinsic scatter of of 0.31~dex for the virial product 
($\propto \Rb \Delta V^2$)  when the line dispersion (in contrast to 
the FWHM) is used. This scatter is equivalent to the intrinsic scatter 
we find for the \Rb--\Rd\ relation ($\lesssim0.25$ dex) which will be the main contributor to the continuum-based virial product.  
We expect the continuum-based method scatter can be reduced 
with future observations of the BLR and the dust continuum size close in time, 
but we discuss the caveats in Section~\ref{ssec:cav}.  
We believe the direct dust continuum size measurement, which we have
shown is tightly linked to the BLR size, is promising to provide high-accuracy BH 
masses in the future (more discussion in Section~\ref{sec:fut}).

\subsection{Caveats}
\label{ssec:cav}

The primary goal of this work is to propose the idea of measuring the BH 
mass based on the dust continuum size, particularly with time-efficient OI 
observations.  The current calibration is not ideal because the BLR 
and dust continuum are not measured in the same AGN luminosity state.  
The line width measurements in the lower part of Table~\ref{tab:smp} are 
collected from different epochs too.  This may introduce significant uncertainty 
on the BH mass (Table~\ref{tab:der}), because the latter is 
$\propto \Delta V^2$.  Due to this reason, we caution that the BH mass and 
Eddington ratio in Table~\ref{tab:der} are only for the purpose to discuss 
the new method instead of rigorous measurements.

In practice, it may not be easy to measure the BLR and dust continuum 
size when they reflect the same luminosity state in order to calibrate 
the \Rb--\Rd\ relation.  It is more feasible to measure the BLR and the dust 
continuum sizes close in time (see also Section~\ref{sec:fut}).  In this way, 
the different time lags of the BLR and the hot dust will contribute to 
the intrinsic scatter of their size relation.  The BLR size and the line width 
may show quicker and stronger variations than the dust continuum because the BLR 
has 5--10 times smaller size than the dust continuum.  We expect a stronger 
averaging effect on the dust continuum too.  For an extreme example, the dust 
continuum size may be correlated with a long-term average of the AGN luminosity 
over the previous several years, as discussed by \cite{Kishimoto2013} for 
NGC~4151.  Future observations are important to quantify how much we can improve 
the \Rb--\Rd\ relation from the first calibration provided in this work. 

The current calibration of the \Rb--\Rd\ relation is limited by 
the sample size.
As briefly discussed in Section~\ref{sec:sc}, various physical mechanisms may 
lead to the deviation of $R \propto L^{0.5}$ for both the BLR and the hot dust.  
Such deviations may reflect variations of BLR and hot dust structures, which may 
not necessarily follow each other.  For example, the BLR radius may have a 
secondary dependence on the BH accretion rate \citep{Du2015,Du2019}, while 
the dust sublimation radius does not depend on it 
\citep{Barvainis1987,Kishimoto2007}.  This  difference means that the \Rb--\Rd\ 
relation may depend on some secondary physical parameters, such as the accretion 
rate.  Such a dependence is not found in the current sample 
(Section~\ref{ssec:dev}), but we cannot rule out the possibility that we are 
limited by the parameter space of the current sample.  Following 
\citet[][Equation~2]{Du2015}, we calculate the dimensionless accretion rate, 
$\mathcal{\dot{M}}$ and find that our targets span 
$10^{-3} \lesssim \mathcal{\dot{M}} \lesssim 200$, with 23\% of our 
targets falling in their super-Eddington regime ($\mathcal{\dot{M}}>3$).  
We do not find a correlation between $\Rb/\Rd$ and $\mathcal{\dot{M}}$, but 
this correlation is worth revisiting with future larger samples including more 
high accretion rate AGNs.

\section{Prospects}
\label{sec:fut}

OI observations have great potential to measure the BLR and continuum of AGNs 
in the future. 
Current GRAVITY observations are limited to the brightest targets ($K<11$).  
With the ongoing upgrade to significantly improve its sensitivity and sky coverage, GRAVITY+ will 
be capable of observing $K<13$ AGNs in on-axis mode or even 
fainter AGNs in off-axis mode with a phase reference source at $\lesssim 30\arcsec$ \citep{GP2022}, 
enabling observations of thousands of AGNs from $z \lesssim 0.2$ out to 
$z \gtrsim 2$.  

Dust continuum sizes are a side product of GRAVITY(+) spectroastrometry 
observations of the BLR if the AGN itself is bright enough to be the phase 
reference.  Half of the AGN emission is split into the low-resolution beam 
combiner (or the `fringe tracker'), which is used as the phase reference of 
the long-time exposure in the science channel to measure the BLR 
spectroastrometric signal \citep{GC2017FL}.  The dust continuum size can be 
measured from the visibility of the fringe tracker data.  
Since the spectroastrometric measurement constrains the BLR geometry and 
dynamics and the BH mass for individual AGNs, we can use the simultaneous 
measurements of the BLR and continuum to calibrate the virial factor in 
Equation~(\ref{eq:vir}) directly for \Rd\ instead of \Rb.  This approach will 
further improve the accuracy of BH masses from the continuum-based method.

GRAVITY can efficiently measure the dust continuum size, e.g. 
$\lesssim 1$~hour observation for one source with current sensitivity 
(GRAVITY Collaboration et~al. in preparation).  We can put all AGN light into 
the fringe tracker so that the continuum method can measure the BH mass of AGNs 
a factor of two (or 0.75~mag) fainter than the spectroastrometry method.  
GRAVITY+ observations will be able to measure the dust continuum size and 
derive the BLR size of a few hundreds of $z\lesssim 0.2$ AGNs, 
including many sources with $\lopt>10^{45}\,\ergs$.  They will be crucial to 
understanding the dependence of BLR properties and the BH accretion on the AGN 
properties.

\section{Summary}
\label{sec:sum}

In this paper, we collect 42 AGNs with dust continuum size measurements 
from RM and/or OI observations.  Among them, the BLR size based on \hb\ RM 
measurements of 26~AGNs are available.  We find close linear relations between 
the BLR and dust continuum radius with an intrinsic scatter of only 0.25~dex.  
The dust continuum radius measured by OI is about twice as large as that 
measured by RM.  Dust continuum radii measured by RM and OI are about five and 
ten times the radius of the BLR, respectively.  We provide simple scaling 
relations to derive the BLR radius based on the dust continuum radius, measured 
with RM and OI separately.  For the remaining 16 AGNs, we calculate 
their BLR radii, BH masses, and Eddington ratios using the \Rb--\Rd\ relations.  
We find that these AGNs consistently follow the BLR \rl\ relation of previous 
RM and GRAVITY measurements.  All targets significantly below 
the \cite{Bentz2013} relation show a high Eddington ratio.  

The accuracy of the continuum-based BH mass is comparable to that of the 
integrated broad emission line RM measurements since the primary uncertainty 
comes from the virial factor.  The primary goal of this paper is to 
propose a new method to measure the BH mass based on the dust continuum size.  
We discuss the caveats of the method in detail.
More continuum observations close in time with BLR measurements will be 
essential to study better the \Rb--\Rd\ relation in the future.  
In particular, it is important to test whether luminous AGNs with large 
BLRs show a different \Rb--\Rd\ relation compared to their low-luminosity 
counterparts.  With its improved sensitivity, GRAVITY+ will be powerful to 
improve the continuum-based method and to efficiently measure the BH 
mass for a large sample of AGNs in the low-redshift Universe using this method.

\begin{acknowledgements}
We thank the anonymous referees for their careful reading and 
suggestions that helped to improve this manuscript.
This research has made use of the NASA/IPAC Extragalactic Database (NED) which 
is operated by the California Institute of Technology, under contract with the 
National Aeronautics and Space Administration.  This research has made use of 
the SIMBAD database, operated at CDS, Strasbourg, France.
\end{acknowledgements}

\bibliography{continuum}{}
\bibliographystyle{aa}

\end{document}